 \pdfoutput=1
\documentclass[traditabstract]{aa} 
\usepackage{graphicx}
\usepackage{natbib}
\usepackage{epsfig}
\usepackage{caption}
\usepackage{afterpage}
\usepackage{float}
\usepackage{amsfonts}
\usepackage{enumerate}
\usepackage{amsmath}
\usepackage{lscape}
\usepackage{txfonts}
\def\vsini{\hbox{$v$\,sin\,$i_\star$}}      

\newcommand{\teff}{$T_{\rm eff}$}

\newcommand{\feh}{[Fe/H]}
\newcommand{\MJ}{{\it M}$_{\rm J}$}

\newcommand{\RJ}{{\it R}$_{\rm J}$}

\newcommand{\ksint}{\texttt{KSint}}
\newcommand{\pastis}{\texttt{PASTIS}}
\newcommand{\ebop}{\texttt{EBOP}}
\newcommand{\corot}{CoRoT}
\newcommand{\amax}{$\alpha_{\mathrm{max}}$}
\newcommand{\tmax}{$t_{\mathrm{max}}$}
\newcommand{\tlife}{$t_{\mathrm{life}}$}
\newcommand{\ingress}{$\mathcal{I}$}
\newcommand{\egress}{$\mathcal{E}$}

\begin{document}

\title{Disentangling planetary and stellar activity features\\ in the CoRoT-2 light curve}

\author{
G. Bruno\inst{\ref{lam}}
\and M. Deleuil\inst{\ref{lam}}
\and J.-M. Almenara\inst{\ref{ipag}} 
\and S. C. C. Barros\inst{\ref{lam},\ref{caup}}
\and A. F. Lanza\inst{\ref{inaf}}
\and M. Montalto\inst{\ref{caup}}
\and I. Boisse\inst{\ref{lam}}
\and A. Santerne\inst{\ref{lam},\ref{caup}}
\and A.-M. Lagrange\inst{\ref{ipag}}
\and N. Meunier\inst{\ref{ipag}}
        }

\institute{
Aix Marseille Universit\'e, CNRS, LAM (Laboratoire d'Astrophysique de Marseille) UMR 7326, 13388, Marseille, France\label{lam}
\and UJF-Grenoble 1/CNRS-INSU, Institut de Plan\'etologie et d’Astrophysique de Grenoble (IPAG) UMR 5274, 38041 Grenoble, France\label{ipag}
\and Instituto de Astrof\'isica e Ci\^{e}ncias do Espa\c co, Universidade do Porto, CAUP, Rua das Estrelas, P-4150-762 Porto, Portugal\label{caup}
\and INAF – Osservatorio Astrofisico di Catania, via S. Sofia 78, 95123 Catania, Italy \label{inaf} 
          }
\date{Received ; accepted }

 
  \abstract
   {
  \textit{Context.} Stellar activity is an important source of systematic errors and uncertainties in the characterization of exoplanets. Most of the techniques used to correct for this activity  focus on an ad hoc data reduction.\\
  \textit{Aims.} We have developed a software for the combined fit of transits and stellar activity features in high-precision long-duration photometry. Our aim is to take advantage of the modelling to derive correct stellar and planetary parameters, even in the case of strong stellar activity.\\ 
  \textit{Methods.} We use an analytic approach to model the light curve. The code \ksint, modified by adding the evolution of active regions, is implemented into our Bayesian modelling package \pastis. The code is then applied to the light curve of \corot-2. The light curve is divided in segments to reduce the number of free parameters needed by the fit. We perform a Markov chain Monte Carlo analysis in two ways. In the first, we perform a global and independent modelling of each segment of the light curve, transits are not normalized and are fitted together with the activity features, and occulted features are taken into account during the transit fit. In the second, we normalize the transits with a model of the non-occulted activity features, and then we apply a standard transit fit, which does not take the occulted features into account.\\
  \textit{Results.} Our model recovers the activity features coverage of the stellar surface and different rotation periods for different features. We find variations in the transit parameters of different segments and show that they are likely due to the division applied to the light curve. Neglecting stellar activity or even only bright spots while normalizing the transits yields a $\sim 1.2\sigma$ larger and $2.3\sigma$ smaller transit depth, respectively. The stellar density also presents up to $2.5\sigma$ differences depending on the normalization technique. Our analysis confirms the inflated radius of the planet ($1.475\pm0.031$ \RJ) found by other authors. We show that bright spots should be taken into account when fitting the transits. If a dominance of dark spots over bright ones is assumed, and a fit
on a lower envelope of the deepest transits is carried out,  overestimating the planet-to-star radius ratio of CoRoT-2 b by almost $3\%$ is likely.\\
   }


\titlerunning{Disentangling planetary and starspot features in the CoRoT-2 light curve}
\authorrunning{G. Bruno et al.}

   \maketitle


\section{Introduction}\label{intro}

Stellar activity is one of the main sources of uncertainty for planet detection and characterization. It causes the emergence of both cooler (therefore darker) and hotter (brighter) than average regions on the stellar photosphere. Such dark and bright spots, or activity features, cross the visible stellar disk as the star rotates, and therefore modulate the amount of flux emitted by the star. These spots are distributed in groups, and vary in size, temperature, and position on the stellar disk along an activity cycle.\\ 
In transit photometry, activity features can induce systematic errors in the determination of the planetary parameters. \cite{czesla2009} showed that transit normalization is affected by non-occulted spots on the stellar disk, which leads to an overestimation of the transit depth. They and \cite{silva-valio2010} discussed how spots occulted during a transit act in the opposite way, producing an underestimation of the same parameter. \cite{leger2009} showed that stellar activity can lead to an underestimation of the stellar density. \cite{csizmadia2013} studied the effect of starspots on the estimate of limb-darkening coefficients. \cite{alonso2009}, \cite{barros2013}, and \cite{oshagh2013} showed that stellar activity can induce apparent transit timing variations (TTV), and can introduce errors in the determination of the transit duration as well.\\ 
Several approaches have been tried to disentangle the signal produced by a planet from the one which comes from activity features. The main attempts focus on the data reduction; that is, data are either corrected for the identified activity signatures, or the activity features are masked.\\
To correct for the errors in the derived planet-to-star radius ratio, \cite{czesla2009} proposed adopting a different transit normalization technique to the standard one. The standard normalization consists in dividing each transit profile by a low-order polynomial fitted to the flux adjacent to both sides of the transit. With the normalization of \cite{czesla2009}, the out-of-transit flux modulations are taken into account. Moreover, they assumed that dark spots are dominant over bright ones, and proposed using the lower envelope of the deepest transits in order to recover a transit profile closer to the true one. With this approach on CoRoT-2 b, they found a $\simeq 3\%$ larger planet-to-star radius ratio than the one reported in the discovery paper \citep{alonso2008} where a standard approach was used.\\ 
Various methods have been developed in order to model stellar activity features. Some are based on analytic models \citep[e.g.][and references therein]{budding1977,dorren1987,kipping2012,montalto2014}, 
while others make use of numerical techniques \citep[e.g.][and references therein]{oshagh2013soapt,dumusque2014}. The fitting techniques also differ, going from the division of the stellar surface in segments on which a $\chi^2$ minimization is performed \citep[e.g.][]{huber2009}, to maximum entropy regularization \citep{rodono1995,colliercameron1997,lanza1998}, and to modified Markov chain Monte Carlo (MCMC) algorithms to sample the spot parameter space \citep{tregloan-reed2015}.\\ 

In this work, we present a method to take into account the imprint of activity features on the transit parameters. We update an existing code for activity features modelling, and implement it into our Bayesian modelling package \pastis. Then, we use it to model the light curve of \corot-2. The clearly visible activity pattern of the star of this system, both outside and inside the transits, has made it a widely studied case in the literature. Our goal is to take advantage of the modelling in order to more consistently determine the transit parameters.\\
The paper is structured as follows: in Section \ref{model} we present our method; in Section \ref{treatment} we describe the test case and the application of our method; in Section \ref{res} and \ref{disc} the results are presented and discussed; and we conclude in Section \ref{conclusions}.

\section{Method}\label{model}
The modelling of activity features is known to be an ill-posed inversion problem because the involved parameters are highly degenerate. To reach convergence in the fit, some of the parameters are kept fixed, or only a part of the data is fitted \citep[e.g.][and references therein]{lanza2009,huber2009}.\\ 
The computation time required by the modelling is another serious inconvenience. Numerical methods define a high-resolution two-dimensional grid of the stellar surface or of its projection onto a plane, and numerically integrate over two coordinates to obtain the light curve. They can deal with a large set of activity features configurations: in principle, they allow for the modelling of any shape for the features, flux profile, and limb-darkening law. The integration, however, is very expensive in terms of computation and time.\\
Analytic models, on the other hand, use analytic formulae to derive a synthetic light curve. They can be order-of-magnitudes faster, but usually require restricting constraints on the parameters to simplify the equations. The main restrictions are that they require simple circular shapes for the activity features, they assume a small size for the spots relative to the stellar surface, and they do not consider the overlapping of the features or of the features with a planet \citep{kipping2012}.\\
Semi-analytic models were employed in some cases in order to overcome these limitations \citep{beky2014}. Alternatively, analytic models which allow for the activity features to overlap, as well as transit modelling, have recently been presented and implemented in freely available codes (e.g. \texttt{KSint}, \citealp{montalto2014}). Their rapidity of execution allows them to be implemented in MCMC algorithms.\\
MCMC methods have already been proven effective in order to find best-fit values, uncertainties, correlations, and degeneracies for the photometric spot modelling problem \citep{croll2006}. Hence, we decided to exploit the additional transit modelling capabilities of an analytic starspot modelling code and the efficiency of MCMC simulations. We implemented the code \ksint\ into the MCMC algorithm used by \pastis\ \citep{diaz2014,santerne2015}. Hereafter, the combination of the codes will be referred to as \ksint\ + \pastis.\\

\ksint\ models a light curve containing both planetary transits and activity features. The transits, modelled with the formalism of \cite{pal2012}, are characterized by the planet-to-star radius ratio $k_r$, orbital period $P_{\mathrm{orb}}$, orbital inclination $i_p$\footnote{The transit shape is degenerate with respect to the stellar hemisphere that the planet covers. In the formalism used by \ksint, inclinations $< 90^{\circ}$ cover the southern hemisphere of the star, and the latitudes of the corresponding occulted activity features are negative.}, eccentricity $e$, planet argument of pericentre $\omega$, and mean anomaly $M$. The star is assigned an inclination angle $i_\star$, a rotation period $P_\star$, a density $\rho_\star$, and quadratic limb-darkening coefficients $u_a$ and $u_b$. The activity features are characterized by the same limb-darkening law as the star.\\
The features are assumed to be spherical caps. Each of them is described by four parameters: longitude $\lambda$, latitude $\phi$, angular size $\alpha$, and contrast $f$. The code models dark and bright spots. It should be noted that bright spots are not faculae, as their contrast does not change from the centre to the limb of the star, as happens for faculae. Bright spots, instead, can be modelled with the same limb-darkening law used for dark spots.\\
In the original version of the code, the time evolution of the features, which has been observed for many stars, is not included. The addition of parameters for feature evolution worsens the problems of correlation, degeneracy, and non-uniqueness of the solution. On the other hand, it allows longer parts of the light curve to be fitted.\\
We therefore introduced a simple law for activity features evolution in \ksint. Following the prescription of \cite{kipping2012}, we used a linear variation of the angular size. The size parameter $\alpha$ was translated into the maximum size reached by a feature during its evolution, \amax. Then, four parameters were added to the description of every feature: 1) the time at which the maximum size is reached, \tmax; 2) the time during which the feature keeps its maximum size, \tlife; 3) the time of growth from zero to maximum size, \ingress; and 4) the time of decay from maximum to zero size, \egress.\\
The features of our model should be considered as representative of groups of features, rather than features taken individually. This allows large sizes and lifetimes to be used without losing physical meaning.\\ 
\pastis\ uses three other parameters to model a light curve: a normalized flux offset, a contamination term, and an instrumental jitter.
 
\section{Application to \corot-2}\label{treatment}

\begin{figure*}[!htb]
\centering
\includegraphics[scale = 0.45]{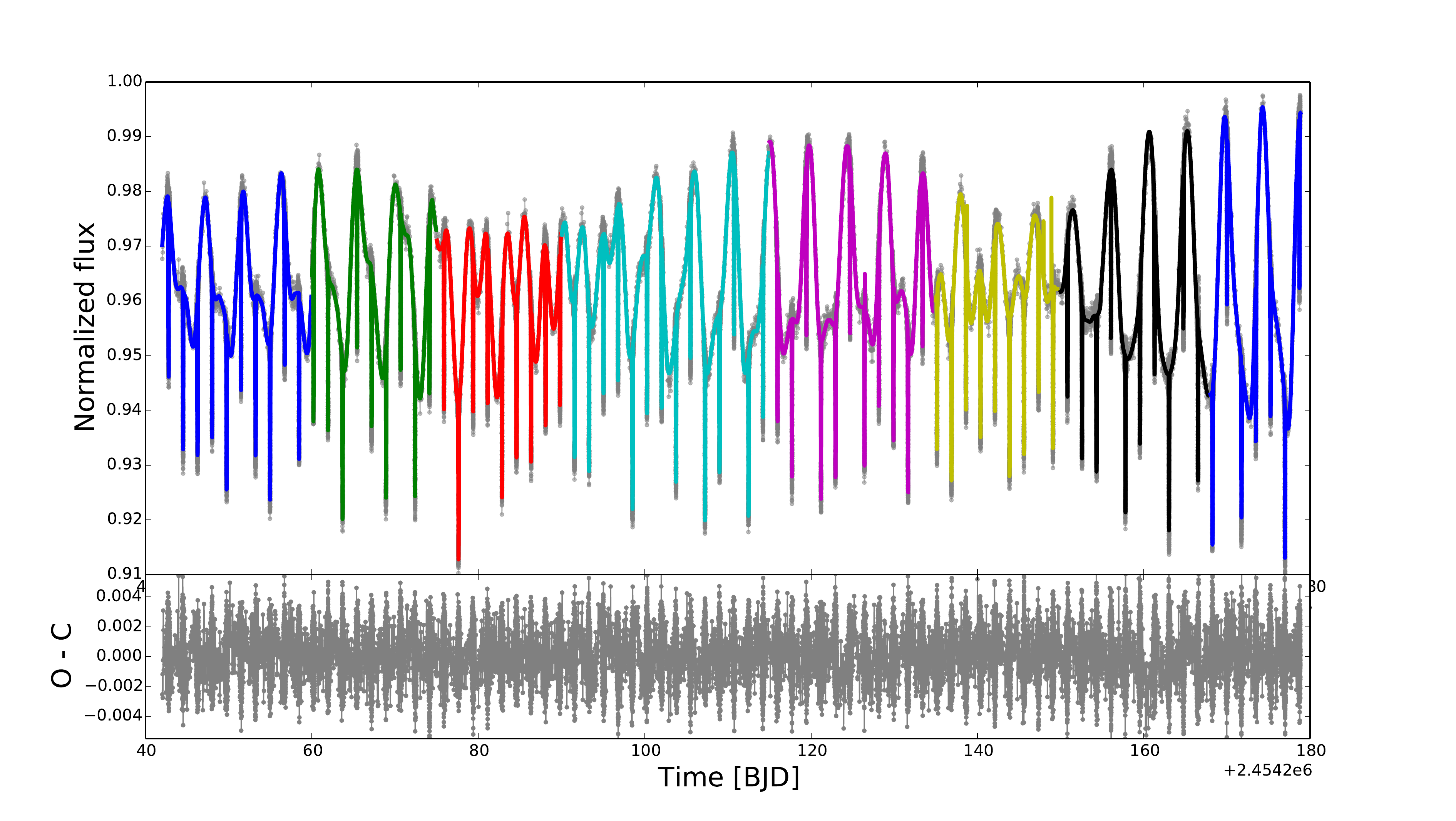}
\caption{Model light curve from \ksint\ plotted over the data. The eight segments of the fit are divided by colour. The residuals are shown in the lower panels; the error bars are not shown for clarity. The larger amplitude of the residuals in correspondence with the transits is due to the full resolution kept for the transits. The out-of-transit binning is of 2016 s, and inside the transits 32 s.}
\label{fit_all}
\end{figure*}

\subsection{Presentation of the system: Previous studies}
\corot-2 A is a G7V-type star observed during the LRc01 run of the \corot\ space telescope. It hosts the hot Jupiter \corot-2 b \citep{alonso2008},  which has mass  $3.31\pm 0.16 $ \MJ\ and radius $1.465 \pm 0.029$ \RJ. The orbit of the planet has a period of 1.74 days, and is almost aligned with the stellar equator \citep{bouchy2008}. Its radius is about 0.3 \RJ\ larger than expected for an irradiated hydrogen-helium planet of this mass. Models strive to explain a longer contraction time during the evolution of the planet, and allow for scenarios with a 30 to 40 Myr old, pre-main sequence host star \citep{guillot2011}. A detailed study of several age indicators favours a main sequence star with an age between 100 and 300 Myr, while the inflated radius of the planet can be explained by a transient tidal circularization and a corresponding tidal heating in the interior of the planet \citep{gillon2010}. However, the system might actually be much older, and the radius anomaly of the planet can be explained by a stellar companion gravitationally bound to the system \citep{schroter2011}. If the star were effectively older than observations suggest, its high level of activity may have been enhanced by its tidal interaction with the planetary companion \citep{poppenhaeger2014}. The main characteristics of the system are listed in Table \ref{presc2}.\\
The light curve, shown in Figure \ref{fit_all}, indicates that a varying fraction of the stellar surface -- up to a
few tens of percent --  is covered by activity features  \citep{lanza2009,huber2009,silva-valio2010}. Fits with a few and with several features have been performed on CoRoT-2, independently from the study of the planet.\\
\cite{wolter2009} worked on a single transit to model an occulted spot. They constrained the size of the spot (between 4.5 and $10.5^\circ$) and its longitude with a precision of about $1^\circ$. \cite{lanza2009} fitted segments of light curve no longer than 3.2 days because the evolution of activity features was not allowed for in their models. They removed the transits and used both a three-spot model and a maximum-entropy regularization method, finding a stellar rotation period of $\sim 4.5$ days. They recovered two active longitudes on different hemispheres, and estimated the relative differential rotation of the star to be lower than $\sim 1\%$. They also measured cyclic oscillations of the total area covered by active regions with a period of $\simeq 29$ days, and found that the contribution of faculae to the optical flux variations is significantly lower than in the present Sun.\\
\cite{frohlich2009} presented a Bayesian analytic spot model of the out-of-transit part of the light curve. They modelled three long-lived spots and recovered the same active longitudes as those found by \cite{lanza2009}. However, their model determines a differential rotation rate which exceeds by one order of magnitude the value found by these authors. \citeauthor{frohlich2009} state that their assumption on spots longevity, not adopted by \citeauthor{lanza2009}, is necessary to recover such a high value of differential rotation.\\ 
\cite{huber2009} modelled the light curve, including transits, over two stellar rotations. They fitted the activity features on the whole light curve but not the transit parameters, and modelled the transits with a combination of the values published by \cite{alonso2008} and \cite{czesla2009}. They managed to reproduce the photometric signal with a  low-resolution surface model of 36 longitude strips, and found that the belt occulted by the planet (close to the stellar equator) is $\sim 6\%$ darker than the average remaining surface. This study was extended to the whole light curve by \cite{huber2010}, whose results are in agreement with the previous study. Significant indications of stellar differential rotation were found.\\
\cite{silva-valio2010} analysed the occulted features inside all transits, but excluded the out-of-transit part of the light curve from the fit. Up to nine spots per transit were needed for the fit. These authors found the size of spots to be between 0.2 and 0.7 planetary radii, and a spot coverage of 10-20\% in the belt transited by the planet. For the spots, they found contrasts between 0.3 and 0.8. Finally, they found a planet-to-star radius ratio of the deepest transit (assumed to be less affected by spots) of 0.172, i.e. 3\% larger than the value found by \cite{alonso2008}.\\
The occulted features were modelled in a spot map for every transit by \cite{silva-valio2011}. For this map, 392 spots were used. The evolution timescale of the coverage of active regions in the transit chord was found to be between 9 and 53 days.\\

All these approaches are complementary and rely on specific assumptions to assess a different aspect of the light curve. However, none takes full advantage of the information encoded in the out-of-transit part of the light curve in order to properly correct the imprint of stellar activity on the transit profile. This was the motivation for our study: with our model, we aim at a more thorough exploration of the impact of the active regions on the transit parameters.

\subsection{Data reduction}\label{phot}
We used the light curve processed by the \corot\ N2 pipeline\footnote{The technical description of the pipeline is available at \texttt{http://idoc-corotn2-public.ias.u-psud.fr/\\jsp/CorotHelp.jsp}.}. Only the part sampled every 32 s was used ($\sim 145$ over 152 days of observation) to take advantage of the full resolution inside transits. The points classified as poor in quality, including those related to the South Atlantic Anomaly, were discarded. The data were first 3$\sigma$-clipped and the transits were identified with the ephemerides given by \cite{alonso2008}.\\
Five data sets were prepared, as summarized in table \ref{tabdatasets}.
\paragraph{Data set $\mathcal{S}$.} This data set was prepared as a reference to be compared to a standard transit analysis. The flux was divided by its median value. A second-order polynomial was fitted to the flux adjacent to the transits. For this, a window as long as a transit duration was considered on the sides of each transit. The flux was then divided by the resulting polynomial. 
\paragraph{Data set $\mathcal{CZ}$.} This data set was prepared for comparison with a standard transit fit, where non-occulted features are taken into account in the data reduction phase, as done by \cite{czesla2009}. The normalization was performed following the prescription of these authors. For every bin $i$, we calculated the normalized flux $z_i$ as
\begin{equation}
z_i = \frac{f_i - n_i}{p} + 1,
\label{czesla}
\end{equation}
where $f_i$ is the observed flux in bin $i$, $n_i$ is the value of the second-order polynomial fitted to the out-of-transit local continuum in the same bin, and $p$ is the largest flux value in the light curve. This last parameter is assumed to be the least affected by dark spots.
\paragraph{Data set $\mathcal{SP}$.} This data set was used for the fit of activity features (the label stands for ``spots'') while fixing the transit parameters. To reduce the computation time, we followed \cite{huber2009} and sampled the out-of-transit flux every 2016 s (33.6 min). This corresponds to about one-third of the orbital period of the CoRoT satellite \citep{auvergne2009}. Given the stellar rotation period, this corresponds to a resolution of $\sim 1.9^{\circ}/$bin, which is sufficient not to significantly affect the out-of-transit flux measurements. The transits were sampled every 160 s, i.e. with a resolution of $\sim 0.1^{\circ}/$bin. The transits are $\sim 0.09$ days long, which means about 49 points were left for each transit. This sampling keeps enough information to force some features to be modelled inside the transits. The uncertainty of the binned data was calculated as the standard deviation of the points in each bin, divided by the square root of the number of points in the binning window. The average relative uncertainty on this data set is $3.9\cdot 10^{-4}$. After resampling, the flux was divided by its maximum value and no transit normalization was performed.
\paragraph{Data set $\mathcal{SPT}$.} This data set was prepared for active regions and transit fitting. It was prepared as data set $\mathcal{SP}$, but full resolution was kept inside the transits. The average relative standard deviation on this data set is of $1.12 \cdot 10^{-3}$.
\paragraph{Data set $\mathcal{MN}$.} The out-of-transit model fitting data set $\mathcal{SP}$ was used to normalize the transits. The preparation of this data set is described in more detail in Section \ref{evolution}.

\begin{table*}[!htb]
\caption{Data sets prepared for the modellings.}
\begin{tabular}{l|c|c|c|c}
\hline\hline
Name & Normalization & Part of the light curve used & Out-of-transit sampling & Inside-transit sampling\\
\hline
$\mathcal{S}$ & Local (as \citealp{alonso2008}) & Transits & ... & 32 s\\
$\mathcal{CZ}$ & Based on flux maximum (as \citealp{czesla2009}) & Transits & ... & 32 s\\
$\mathcal{SP}$ & No & All & 2016 s & 160 s\\
$\mathcal{SPT}$ & No & All & 2016 s & 32 s\\
$\mathcal{MN}$ & With out-of-transit model & Transits & ... & 32 s\\
\hline
\end{tabular}
\label{tabdatasets}
\end{table*}

\subsection{Modelling approaches}\label{evolution}
We performed two reference fits with standard approaches. Then, we performed three fits including activity modelling.

\subsubsection{Reference fits}

\paragraph{Fit $\mathcal{S}$: activity neglected.}\label{fital}
Fit $\mathcal{S}$ was performed on data set $\mathcal{S}$ as a standard transit fit. We used a modified version of the \ebop\ code\footnote{Hereafter, for simplicity, \ebop.} \citep{nelson1972,etzel1981,popper1981} included in the MCMC functionality of \pastis. The priors for the free parameters are listed in table \ref{tabpriors}. Because of the biases in our current knowledge of the radius distribution of planets, the incompleteness of the transit surveys, and other systematic effects, we followed \cite{diaz2014} and used a Jeffreys prior for the planet-to-star radius ratio $k_r$. We used a Jeffreys prior for the orbital separation-to-stellar radius ratio $a/R_\star$, which is used by \ebop\ in place of the stellar density. A normal prior was also used for the orbital period $P_{\mathrm{orb}}$. To impose an isotropic distribution on orbit orientations, we set a sine prior for the orbit inclination $i_p$. Uniform priors were used for the linear and the quadratic limb-darkening coefficients $u_a$ and $u_b$. Following \cite{alonso2008}, we fixed $e=0$, and consequently also $\omega=0$. The flux offset and the jitter were fitted on the whole light curve.

\begin{table*}[!tbhp]
\caption{\label{tabpriors} Prior distributions used in the combined fit with \ksint\ + \pastis. $\mathcal{U}(a, b)$ stands for a uniform distribution between $a$ and $b$; $\mathcal{N}(\mu, \sigma)$ indicates a normal distribution with mean $\mu$ and standard deviation $\sigma$; 
$\mathcal{S}(a, b)$ represents a sine distribution between $a$ and $b$; finally, $\mathcal{J}(a, b)$ means a Jeffreys distribution between $a$ and $b$. $BV$ indicates the value of the maximum-likelihood solution of fit $\mathcal{SP}$.}
\centering
\begin{tabular}{llll}
\hline\hline
\\
& \textit{fit $\mathcal{S}$, $\mathcal{CZ}$, $\mathcal{MN}$} & \textit{fit $\mathcal{SP}$} & \textit{fit $\mathcal{SPT}$} \\
\multicolumn{2}{l}{\emph{Stellar parameters}} \smallskip\\
Stellar axis inclination $i_\star$ [deg]& ... & 87.84 (fixed) & 87.84 (fixed) \\
Stellar rotation period $P_\star$ [days] & ... &  $\mathcal{U}(4.3, 4.7)$ &  $\mathcal{U}(4.3, 4.7)^{(b)}$ \\
\\
\multicolumn{2}{l}{\emph{Parameters of active regions}} \smallskip\\
Longitude $\lambda$ [deg] & ... & $\mathcal{U}(0, 360)$ & $\mathcal{N}(BV, 5)$\\
Latitude $\phi$ [deg]& ... & $\mathcal{U}(-90, 90)$ & $BV$ (fixed)\\
Maximum size \amax\ [deg] & ... & $\mathcal{U}(0, 30)$ & $\mathcal{U}(0, 30)^{(b)}$\\ 
Contrast $f$ & ... &  $\mathcal{U}(0.3, 1.3)$ & $BV$ (fixed) \\
Time of maximum size \tmax$^{(a)}$ [days] & ... & $\mathcal{U}(-10, 50)$ & $BV$ (fixed)\\
Life time at maximum size \tlife\ [days]& ... & $\mathcal{U}(0, 50)$ & $BV$ (fixed) \\
Time of growth \ingress\ [days] & ... & $ \mathcal{U}(0, 50)$ & $BV$ (fixed)\\
Time of decay \egress\ [days] & ... &  $\mathcal{U}(0, 50) $ & $BV$ (fixed)\\
\\     
\multicolumn{2}{l}{\emph{Transit parameters}} \smallskip\\
Radius ratio $k_r$   & $\mathcal{J}(0.14, 0.19)$ &  0.1667 (fixed) &$\mathcal{J}(0.14, 0.19)$\\
Stellar density $\rho_{\star}$ [g cm$^{-3}$]   & ... &1.87 (fixed) &  $ \mathcal{N}(1.87, 0.5)$\\
Semi-major axis to stellar \\
radius ratio  $a/R_\star$ & $\mathcal{J}(5.0,8.0)$ & ... & ... \\
Orbital inclination $i_p$ [deg]   & $\mathcal{S}(80, 90)$ &  87.84 (fixed) &$\mathcal{S}(80, 90)$ \\
Linear limb-darkening coefficient $u_a$  & $\mathcal{U}(0.0, 1.0)$ & 0.41 (fixed)  & $\mathcal{N}(0.41, 0.03)$\\
Quadratic limb-darkening coefficient $u_b$  & $\mathcal{U}(0.0, 1.0)$ & 0.06 (fixed)  & $\mathcal{N}(0.06, 0.03)$\\
Orbital period $P_{\mathrm{orb}}$ [days]  & $\mathcal{N}( 1.7429964, 1.0\cdot 10^{-5})$ & 1.7429964  (fixed) & $\mathcal{N}( 1.7429964, 1.0\cdot 10^{-5})$ \\
Orbital eccentricity $e$   & 0 (fixed)  & 0 (fixed) & 0 (fixed) \\
Argument of periastron $\omega$ [deg]   & 0 (fixed)    & 0 (fixed) & 0 (fixed)  \\
\\
\multicolumn{2}{l}{\emph{Instrumental parameters}} \smallskip\\
Flux relative offset & $\mathcal{U}(0.99, 1.01)$ &  $\mathcal{U}(0.97, 1.01)$ & $\mathcal{U}(0.97, 1.01)^{(b)}$\\
Flux jitter [ppm]        & $\mathcal{U}(0, 0.01)$ & $\mathcal{U}(0, 0.01)$ & $\mathcal{U}(0, 0.01)$\\
Contamination  [\%] & $\mathcal{N}(8.81, 0.89)$ & 8.81 (fixed) &  8.81 (fixed)  \\
\\
\hline
\end{tabular}
\begin{list}{}{}
\item \small$^{(a)}$ Referred to the initial time of a segment. $^{(b)}$ Starting from the best solution of fit $\mathcal{SP}$. 
\end{list}
\end{table*}

\cite{alonso2008} used a $5.6 \pm 0.3 \%$ contamination rate for CoRoT-2, as calculated before the CoRoT launch using a set of generic PSFs \citep[contamination level 0,][]{llebaria2006}. \cite{gardes2012} updated this value to $8.81 \pm 0.89\%$ using a more realistic estimate of the CoRoT PSF (contamination level 1). We adopted this last value, and fitted the contamination rate using this prior. A larger contamination rate increases the variability amplitude of the star and therefore of the measured transit depth. We expect this latter to increase on average by $(1-0.056)/(1-0.081) = 2.7\%$.\\
The same MCMC simulation was also performed with \ksint\ instead of \ebop\ in order to verify its correct behaviour for a standard transit-only fit.\\
Ten chains were run for every MCMC set. The chains were then thinned according to their correlation length and merged into a single chain, giving the credible intervals for the parameters of that segment. To be considered robust, a merged chain was required to consist of at least a thousand uncorrelated points.

\paragraph{Fit $\mathcal{CZ}$: activity taken into account during normalization.}
For this fit we used the same method as for fit $\mathcal{S}$, but on data set $\mathcal{CZ}$.

\subsubsection{Fits that include activity}
Because of the high level of degeneracy and correlation of the activity features parameters in the starspots modelling problem, different Markov chains do not converge to the same values for the parameters. We therefore divided the fit into two steps. First, only the parameters of the active regions were fitted with a MCMC run. This resulted in a local maximum-likelihood solution. Starting from this solution, we then used another MCMC run to sample the posterior distribution of a subset of key activity features parameters together with the transit parameters. Eventually, we performed a transit fit by taking advantage of the maximum-likelihood solution for the activity features.

\paragraph{Fit of active regions (Fit $\mathcal{SP}$).}\label{spotsonly}
This fit takes into account the activity features only, and prepares the fit of the transits. The light curve modelling was carried out on the data set $\mathcal{SP}$. The transit parameters ($P_{\mathrm{orb}}, \, \rho_\star, \, i_p, \, k_r, \, e, \, \omega, \, u_a, \, u_b$) were fixed to the values of \cite{alonso2008}. A uniform prior was imposed for the mean anomaly $M$. The priors indicated in table \ref{tabpriors} were used. The quasi-perpendicularity between the sky-projected stellar spin axis and the planetary orbit found by \cite{bouchy2008} allowed us to model the system as if these two were perpendicular. \cite{alonso2008} found the planet inclination to be of $87.84^\circ$ with respect to the plane of the sky. We therefore fixed the stellar inclination to the same value with respect to the line of sight. A uniform prior centred on the value found by \cite{lanza2009} was set for the stellar period $P_\star$. Uniform priors were used for all the parameters of the activity features. For each feature, \amax\ was limited between 0 and $30^\circ$. The largest size was found to be sufficient for the modelling.\\
The analysis of \cite{lanza2009} showed that bright regions have a minor impact on the light curve of CoRoT-2 compared to dark features. However, these authors obtained their result after removing the transits from the data set. Instead, because our data set includes transits, we decided to model both dark and bright spots. We set a prior for the contrast $f$ between 0.3 and 1.3. These values represent a wider interval than the one defined by the typical sunspot bolometric contrast of 0.67 \citep{sofia1982,lanza2004} and the bolometric facular contrast 1.115 adopted by \cite{foukal1991} and \cite{lanza2004}. Latitudes were left free between $-90^\circ$ and $90^\circ$ in order to allow both non-occulted and occulted spots to be modelled.\\
For each feature, a uniform prior was set for \tmax. Values of \tmax\ external to the time limits of the light curve were allowed in order to induce the fit to exclude possible features in excess. The MCMC could do this by adjusting \tmax, \ingress, and \egress, to which uniform priors were assigned, too.\\ 
The contamination factor was fixed to the value found by \cite{gardes2012} (8.81\%). This parameter was fixed because active regions can be considered a contaminant factor \citep{csizmadia2013}. A uniform prior was set for the flux relative offset and the jitter.\\

The optimal number of active regions was determined by trials. We increased it until we obtained residuals with a normally distributed dispersion centred on zero and with a width comparable to the photometric data dispersion. Our model needs several tens of active regions to model the entire light curve, in agreement with the results of \cite{silva-valio2010} and \cite{silva-valio2011}. This implies hundreds of free parameters, which our computer cluster is not able to handle. Therefore, the light curve was divided into shorter parts which need the modelling of fewer features. The initial and ending time of these segments are indicated in table \ref{tsegments}. The segments have a duration of $\sim15-25$ days (four to six stellar rotations), and can be fitted with six to nine evolving features. Longer segments tend to produce worse fits. In Figure \ref{fit_all}, the segments are highlighted in colour. It can be seen how the segments are related to the different phases of activity in the light curve. The brightness variations grow in amplitude, reach a maximum, and shrink again. The duration of the segments is consistent with the lifetime of individual spots and active regions found by \cite{lanza2009}, i.e. between 20 and 50 days.\\
To connect the solutions of consecutive segments, the evolving features with non-zero size at the end of a given segment were kept for the initial values of the features of the next segment. Up to eight features were kept from one segment to the next. For such features, we used the same latitude $\phi$, and constrained the longitude $\lambda$ around the value of the solution in the previous segment. For this parameter, we chose a normal prior with standard deviation equal to $10^\circ$, and constrained $c$ in order to force the feature to remain darker or brighter than the stellar surface. The parameter \amax\ was again set as a free parameter with a uniform prior, starting from the maximum-likelihood value of the previous segment. The evolution parameters were left free in order to allow for the feature size to grow after a decaying phase and vice versa. Other activity features were added to those kept to connect the segments using the same priors described in table \ref{tabpriors}. In this way, the appearance of new features along the light curve were modelled. The total number of features per segment (fixed and free) is between six and nine.\\
The flux offset was fitted separately in every segment, starting the chains of each segment from the maximum-likelihood value of the previous one. In this way, a possible photometric long-term trend was taken into account.\\
Ten chains of up to $3 \cdot 10^5$ iterations were run for each segment. Most of the chains of each segment were observed to reach the end of their burn-in phase with this number of iterations. We considered a  chain to have reached a steady state if, after its burn-in phase, its likelihood did not show significant increments for a few thousand steps. We then extracted the maximum-likelihood solution among the chains satisfying this criterion.\\
The probability of finding the global maximum of the likelihood function increases with the number of chains run. However, we note that our method does not allow us to identify $the$ global solution of the starspot inversion problem. With this approach we obtained a fit of the activity features parameters suitable for the fit of the transit parameters. Also, the maximum-likelihood solution allowed us to explore the active regions configuration on the stellar surface. This is discussed in greater detail in the following sections.

\paragraph{Spot-transit fit (fit $\mathcal{SPT}$).}
We carried out a simultaneous fit of the transit parameters and of the activity features both in-transit and out-of-transit. This fit was performed on data set $\mathcal{SPT}$, divided into segments as for fit $\mathcal{SP}$. This time the transit parameters were set as free parameters, while the parameters of the active regions were fixed to the maximum-likelihood solution of fit $\mathcal{SP}$ for each segment. An exception was made for the size \amax\ and the longitude $\lambda$ of the activity features. The parameter \amax\ affects primarily the planet-to-star radius ratio. Its fit was started from the value of the best solution of fit $\mathcal{SP}$. The longitude affects the position of a feature in the transit profile, with consequences for the transit duration, and therefore the limb-darkening coefficients, $i_p$, and $\rho_\star$. We used a normal prior with standard deviation equal to $5^\circ$, centred on the best likelihood value.\\
The priors of the transit parameters are the same as for fit $\mathcal{S}$ and $\mathcal{CZ}$. Two parameters were set differently. Instead of $a/R_\star$, \ksint\ uses $\rho_\star$, for which we used a Gaussian prior centred on the \cite{alonso2008} value and a standard deviation $0.35 \, \rho_\odot$. Normal priors were set for the limb-darkening coefficients, using the results of \cite{alonso2008}.\\
In  this phase, ten chains from 1.5 to $2\cdot 10^5$ steps were employed to reach convergence. For each segment, the chains were then thinned according to their correlation length and merged into a single one. The results and the uncertainties for each segment were obtained from the respective merged chains, as for fit $\mathcal{S}$ and $\mathcal{CZ}$.\\
This fit resulted in slightly scattered transit parameters among the various segments, as discussed in Section \ref{joint}. The posterior distributions allowed us to explore the impact of bright spots on the deepest transit and the correlations between activity features and transit parameters. This will be discussed in Section \ref{disc}.

\paragraph{Model-based normalization (fit $\mathcal{MN}$).}
The fit consists of a standard transit fit  where transit normalization is performed via the modelling of the out-of-transit features. The maximum-likelihood solutions obtained from fit $\mathcal{SP}$ were re-computed without the planet. They were merged into a single light curve and used to normalize the transits, obtaining data set $\mathcal{MN}$ (Section \ref{phot}). Equation \ref{czesla} was used, with $n_i$ indicating the model and $p$ the flux offset value. For each segment, this parameter was fixed to the maximum-likelihood value obtained with fit $\mathcal{SP}$. Its uncertainty, obtained with fit $\mathcal{SPT}$, was quadratically added to the standard deviation of the flux. In this way, the uncertainty due to the normalization technique was evaluated.\\
Once the transits were normalized, a standard transit fit was performed with \ebop. This code was used because of its higher computation speed compared to \ksint. The priors are the same as fit $\mathcal{S}$ and $\mathcal{CZ}$. The contamination was fitted as well because in-transit activity features were no longer modelled.

\section{Results}\label{res}

Figure \ref{fit_all} presents the best model of fit $\mathcal{SPT}$ plotted over the light curve. The segments are divided by colour. The parameters of the active regions (yielded by fit $\mathcal{SP}$) and the transit parameters (fit $\mathcal{SPT}$) are discussed separately.

\subsection{Parameters of the active regions}\label{sppar}
\begin{figure}[!htbp]
\centering
\includegraphics[scale = 0.45]{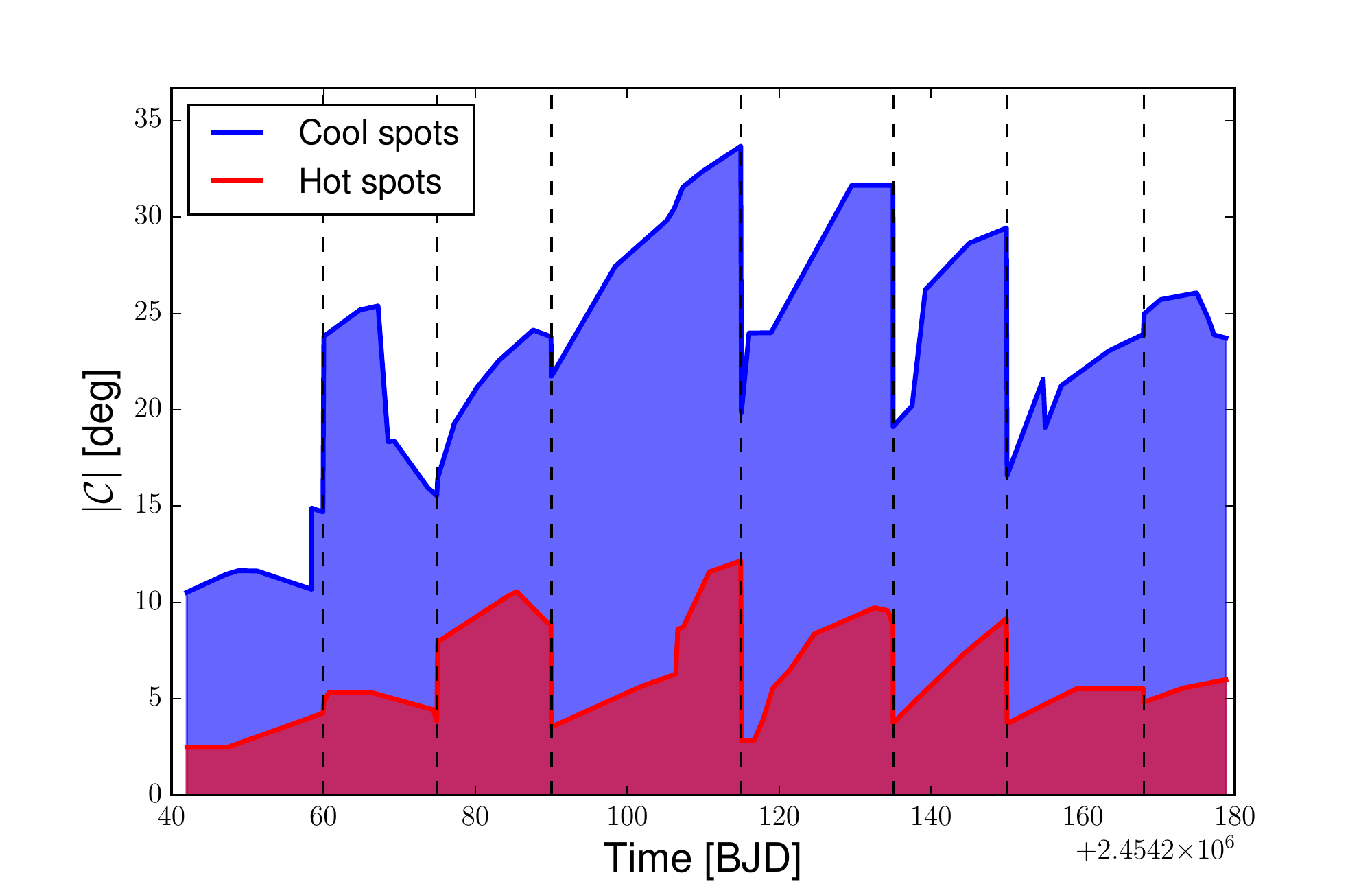}
\caption{Absolute value of the effective coverage factor as a function of time. The vertical dashed lines mark the separations between segments.}
\label{aeff_slice}
\end{figure}

The maximum-likelihood solutions of fit $\mathcal{SP}$ allowed us to recover some properties of the starspot surface coverage for each segment of the light curve. 
We do not expect our solutions to be in full agreement with those presented by other authors because our fit 1) includes the fit of activity features inside transits, which provides a strong constraint on their longitudes; 2) uses both dark and bright spots, which obey the same limb-darkening law; and 3) models the evolution of the features. In alternative modellings, instead, the authors drop one of these points. \\
We computed the effective coverage factor of the stellar surface as a function of the segment of light curve, separately for dark and bright spots. For each segment, this is defined as 
\begin{equation}
\mathcal{C} =\sum_i \alpha_{\mathrm{max},i} (1 - f_i),
\end{equation}
where the sum is run over all the features $i$ with maximum area $\alpha_{\mathrm{max}}$ and contrast $f$. As only the maximum-likelihood solutions were used, no error estimate is possible. A bright spot has $f > 1$, hence yields a negative $\mathcal{C}$.\\
In Figure \ref{aeff_slice}, the absolute value $\mathcal{|C|}$ for dark (in blue) and bright (in red) spots is plotted. It can be noticed that a larger $\mathcal{|C|}$ for dark spots is accompanied by a somewhat larger $\mathcal{|C|}$ for bright spots. \cite{lanza2009} found a faculae-to-dark spots surface ratio between 1 and 2.5 in active regions. We chose not to directly compare our results with theirs because, unlike these authors, we used the contrast of the activity features as a free parameter.\\
Discontinuities of $\mathcal{|C|}$ can be observed on the separations between segments (vertical dashed lines). These is due to the lack of constraints on the features evolution parameters \tlife, \tmax, \ingress, and \egress\ between consecutive segments.  

\begin{figure}[!htbp]
\centering
\includegraphics[scale = 0.45]{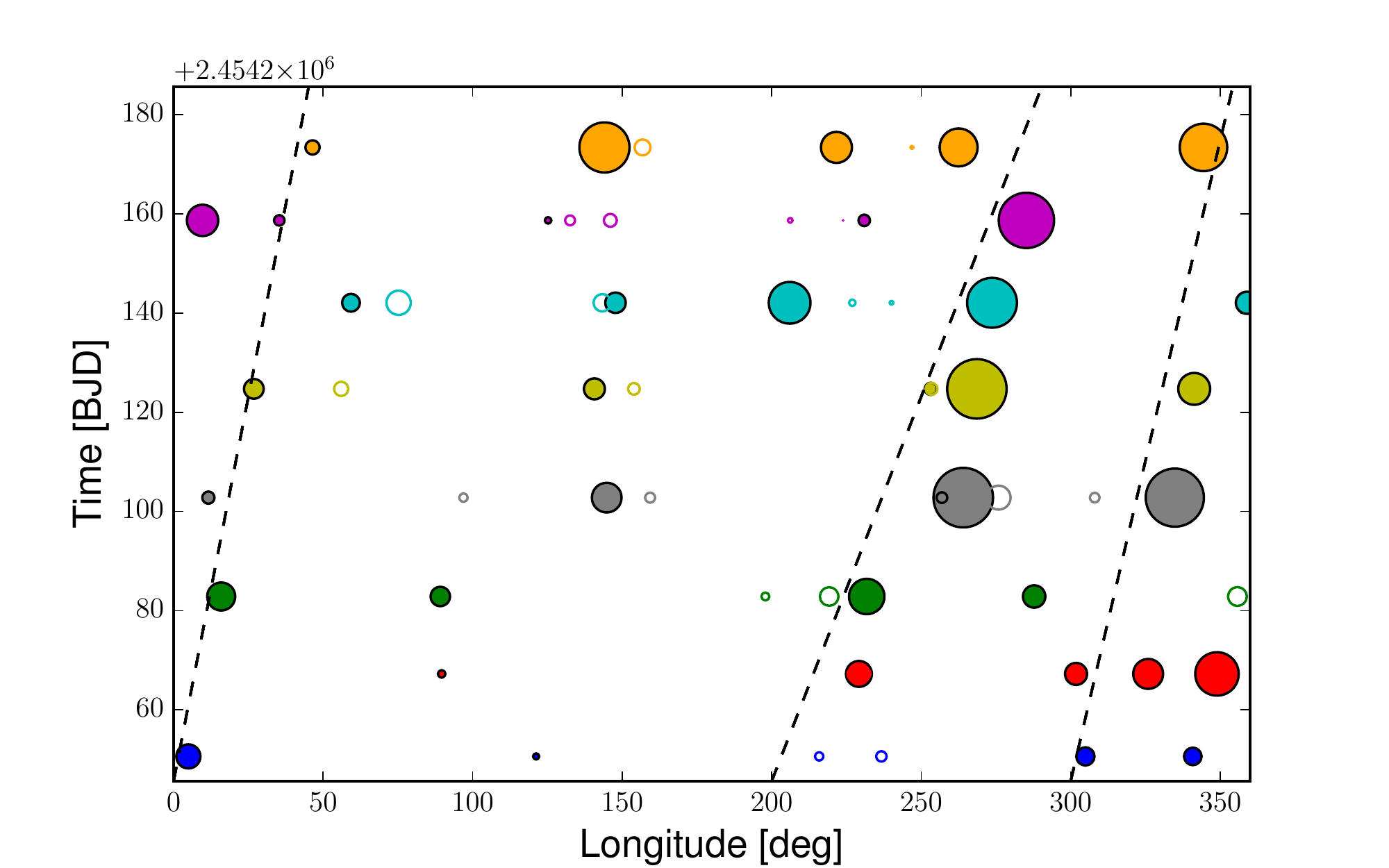}
\caption{Distribution of the activity features as a function of their longitude and time. No latitudinal information is reported. Filled circles represent dark spots, empty circles represent bright ones. Different segments of the light curve are represented by circles of different colour. The relative size of the features compared to the stellar surface was increased for the sake of illustration. Tentative reconstruction of longitudinal migration are represented by diagonal dashed lines.}
\label{aeff_long}
\end{figure}

Figure \ref{aeff_long} represents the activity features for every segment as a function of their longitude ($x$-axis) and time ($y$-axis). No latitudinal information is reported. Filled circles represent dark spots, empty circles represent bright ones. Different colours are used for features in different segments. The sizes of the circles indicate the $\alpha_{\mathrm{max}}$ of the activity features. The relative size compared to the stellar surface was increased for the purpose of illustration. As the maximum size is represented and not its evolution along the light curve, the plot gives an upper-limit coverage of the surface for each segment. Each feature is placed at its respective segment's midtime for clarity. \\
This plot suggests that the activity features are not uniformly distributed along the longitudes. Also, a longitudinal migration towards higher longitudes can be seen for the activity features with $\lambda$ between 0 and $\sim 50^\circ$ and those between 200 and $360^\circ$. Tentative patterns of migration are shown with diagonal dashed lines. Features between 100 and $200^\circ$, instead, do not show the same migration pattern, but keep the rotational period assigned to the star. We argue that assigning different rotation periods to the features, as \cite{frohlich2009} did, would allow longer segments of the light curve to be modelled.\\
The inhomogeneous distribution of the features and their size along the stellar longitudes, shown in Figure \ref{aeff_long}, suggest the presence of differently active longitudes. By using a continuous distribution of active regions with fixed contrast or only three dark spots on the stellar surface, respectively, \cite{lanza2009} and \cite{frohlich2009} achieved a similar result. \cite{huber2010} also confirmed such findings by using a different method. Given our different modelling of the evolution of the features, as well as the degeneracy between their number, their size, and their contrast, we chose not to compare in detail our results with theirs.

Most of the features have \amax $> 5^\circ$. Moreover, the average size of the features does not change among different activity phases (increasing brightness variations, maximum, decrease of the brightness variations; see Section \ref{spotsonly} and Figure \ref{fit_all}).\\
Bright spots are present in every segment. Moreover, in every segment, one or two features (either dark or bright spots) are found to cross the transit chord. As six to nine features are present in each segment, occulted features are a minority. Finally, no preferred values are found for the fitted evolution times \tmax, \tlife, \ingress, and \egress.

\subsection{Transit parameters}\label{joint}
\begin{figure}[!htbp]
\centering
\includegraphics[scale = 0.45]{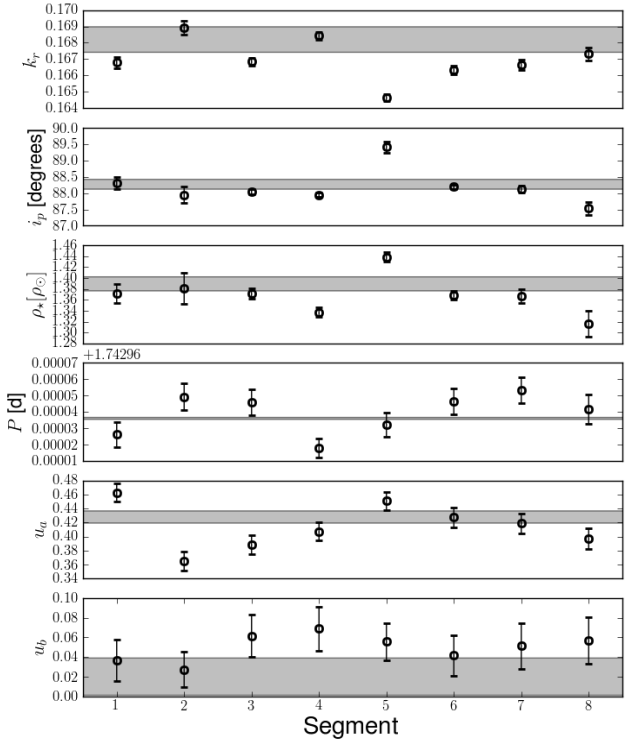}
\caption{From top to bottom: Values and uncertainties of $k_r$, $i_p$, $\rho_\star$, $P_{\mathrm{orb}}, u_a$, and $u_b$ for fit $\mathcal{SPT}$, indicated as a function of the segment of the light curve are in black. The results from fit $\mathcal{MN}$ with their 68.3\% credible intervals are shaded in grey.}
\label{all_post}
\end{figure}

Table \ref{tabastronorm} reports the transit parameters of all fits. The results of fit $\mathcal{SPT}$ were obtained as the average of all the results obtained on each segment. The values obtained for each segment are reported in table \ref{tsegments}. For some segments, the results of fit $\mathcal{SPT}$ are in poor agreement one to each other. This is likely due to statistical fluctuations introduced by the division of the light curve in segments, as will be discussed in more detail in Section \ref{mostrel}. Because of this, the error bars on the average results of fit $\mathcal{SPT}$, in table \ref{tabastronorm}, are larger than on the other fits, whose uncertainties are similar to those found by \cite{alonso2008}. For the same reason, the average results of fit $\mathcal{SPT}$ are only indicated for comparison to the other results, but are not discussed further.
Figure \ref{all_post} shows the results on each segment. The shaded regions correspond to the results of fit $\mathcal{MN}$.\\  

\begin{table*}[!hbt]
\begin{center}{
\caption{\label{tabastronorm} Transit parameters with their 68.3\% credible intervals. Results on the classic fit (fit $\mathcal{S}$), on the light curve normalized as \cite{czesla2009} (fit $\mathcal{CZ}$), on the segments (fits $\mathcal{SP}$ and $\mathcal{SPT}$), and on the model-normalized light curve (fit $\mathcal{MN}$). The stellar density of fit $\mathcal{S}$, $\mathcal{CZ}$, and $\mathcal{MN}$ is derived from the respective $a/R_\star$.}
\begin{tabular}{lllll}
\hline
\hline
\\
\textit{Parameter}  & \textit{fit $\mathcal{S}$} & \textit{fit $\mathcal{CZ}$} &  \textit{fit $\mathcal{SP}$-$\mathcal{SPT}$} & \textit{fit $\mathcal{MN}$} \smallskip\\

$P_{\mathrm{orb}}$ [d]  & $1.74299628 \pm 5.8 \cdot 10^{-7}$ & $1.74299620 \pm 5.9 \cdot 10^{-7}$ & $1.742999\pm 1.2 \cdot 10^{-5}$ & $1.74299609 \pm 5.9\cdot 10^{-7}$\\
$T_0$ [BJD]   & $ 2454237.535398 \pm 2.9 \cdot 10^{-5}$ & $2454237.535399 \pm 2.8 \cdot 10^{-5}$ &  ... & $2454237.535401 \pm 2.8 \cdot 10^{-5}$\\

$i_p$ [degrees]  & $88.34 \pm 0.16$ &  $88.01 \pm 0.13$  & $88.19 \pm 0.51$ & $88.27 \pm 0.15$\\
$k_r$  & $0.16917 \pm 8.5\cdot 10^{-4} $ &  $0.16632\pm  8.0\cdot 10^{-4}$ & $0.1670 \pm 1.2\cdot 10^{-3}$ & $ 0.16820\pm 7.8 \cdot 10^{-4}$\\
$\rho_\star$ [$\rho_\odot$] & $1.395 \pm 0.014$ & $1.361 \pm 0.013$ & $1.369 \pm 0.033 $ & $1.390 \pm 0.013$ \\
$u_a$  & $0.428^{+0.006}_{-0.009}$ & $0.426^{+0.007}_{-0.011}$ & $0.415 \pm 0.030$ & $ 0.428^{0.007}_{-0.011}$ \\
$u_b$ & $0.017^{+0.020}_{-0.013}$ & $0.023\pm 0.023$ &  $0.050 \pm 0.013$ & $0.020^{+0.023}_{-0.016}$ \\
Flux offset  & $1.0000473 \pm 9.0\cdot 10^{-6}$ & $1.0000459 \pm 8.6\cdot 10^{-6}$  & $0.9755\pm 4.1 \cdot 10^{-3}$  & $1.0000467 \pm  9.3\cdot 10^{-6}$\\
Flux jitter  & $0.0009171 \pm 9.1\cdot 10^{-6}$ &  $0.0009283 \pm  8.3 \cdot 10^{-6}$ & ... & $0.0009387 \pm 8.9\cdot 10^{-6}$\\
Contamination [\%]  & $8.92 \pm 0.90$ & $8.81\pm 0.90 $ & ... &$8.83 \pm 0.85$ \\
\\
 \hline

\end{tabular}
}\end{center}
\end{table*}

Differences were found in the planet-to-star radius ratio $k_r$ among all fits carried out on the whole light curve ($\mathcal{S}, \, \mathcal{CZ},$ and $\mathcal{MN}$). This highlights the dependence on the normalization technique which was used. As expected, the parameter $k_r$ is larger if a standard normalization is performed (fit $\mathcal{S}$, $k_r = 0.16917 \pm 8.5\cdot 10^{-4} $). In fact, in this case, non-occulted spots are neglected and cause an overestimate of this parameter. On the other hand, the \cite{czesla2009} normalization (fit $\mathcal{CZ}$) produces a smaller $k_r$ ($0.16632\pm  8.0\cdot 10^{-4}$). This occurs because of the division of the transits profiles by the maximum flux value ($p$), which is assumed to be the least affected by activity. Without a model of activity features, no uncertainty is available for this value. This is the main limitation of this normalization.\\
Fit $\mathcal{MN}$ models non-occulted activity features, therefore, it is not affected by the same issues as fits $\mathcal{S}$ and $\mathcal{CZ}$. It yields a $\sim 1.2\sigma$ smaller $k_r$ value than that obtained with fit $\mathcal{S}$, and $2.3\sigma$ larger than  obtained with fit $\mathcal{CZ}$.\\ 
The $k_r$ on segment 5 is lower than those estimated from the other segments, and higher for $i_p$ and $\rho_\star$. The large contribution of bright spots with respect to dark ones, observed in Figure \ref{aeff_slice} for this segment, suggests that our correction of the transit profiles for bright spots can be refined.\\

The stellar densities $\rho_\star$ and the orbital inclinations $i_p$ are in agreement between fit $\mathcal{S}$ and $\mathcal{MN}$, while they are at less than $2\sigma$ agreement between fit $\mathcal{CZ}$ and $\mathcal{MN}$. Except for segments 5 and 8, the results of fit $\mathcal{SPT}$ for $i_p$ and $\rho_\star$ are consistent with each other.\\ 

The orbital periods $P_{\mathrm{orb}}$ found with all fits on the whole light curve are between $\sim1$ and $3\sigma$ agreement. We note, however, that their measure is scattered among the segments of fit $\mathcal{SPT}$. This can be explained by statistical fluctuations due to number of transits -- no more than about ten -- in  each segment . 
We exclude the effect of spurious activity-induced TTV such as those recovered by \cite{alonso2009} in the CoRoT-2 light curve. These authors found a 7.45 d peak in the periodogram of the residuals of the transit midpoints and attributed it to activity features occulted by the planet. The amplitude of the peak they found is 20 s, while the variations in Figure \ref{all_post} are about one order of magnitude smaller.

The limb-darkening coefficients $u_a$ and $u_b$ were found to be compatible for all fits on the whole light curve. In fit $\mathcal{SPT}$, instead, $u_a$ results in a poorer agreement among the segments despite the imposed tight normal prior. An explanation can be the difficulty of \pastis\ + \ksint\ to recover the limb-darkening coefficients if only a few transits are available as too little information is available to disentangle the effect of activity features on the limb-darkening coefficients. A poorly fitted $P_{\mathrm{orb}}$ could contribute to this problem as well. Another cause could be that the limb-darkening coefficients actually change as a function of the varying coverage of the stellar surface by active regions \citep{csizmadia2013}. This possibility is discussed in Section \ref{mostrel}.

\begin{figure*}[!htbp]
\centering
\includegraphics[scale = 0.4]{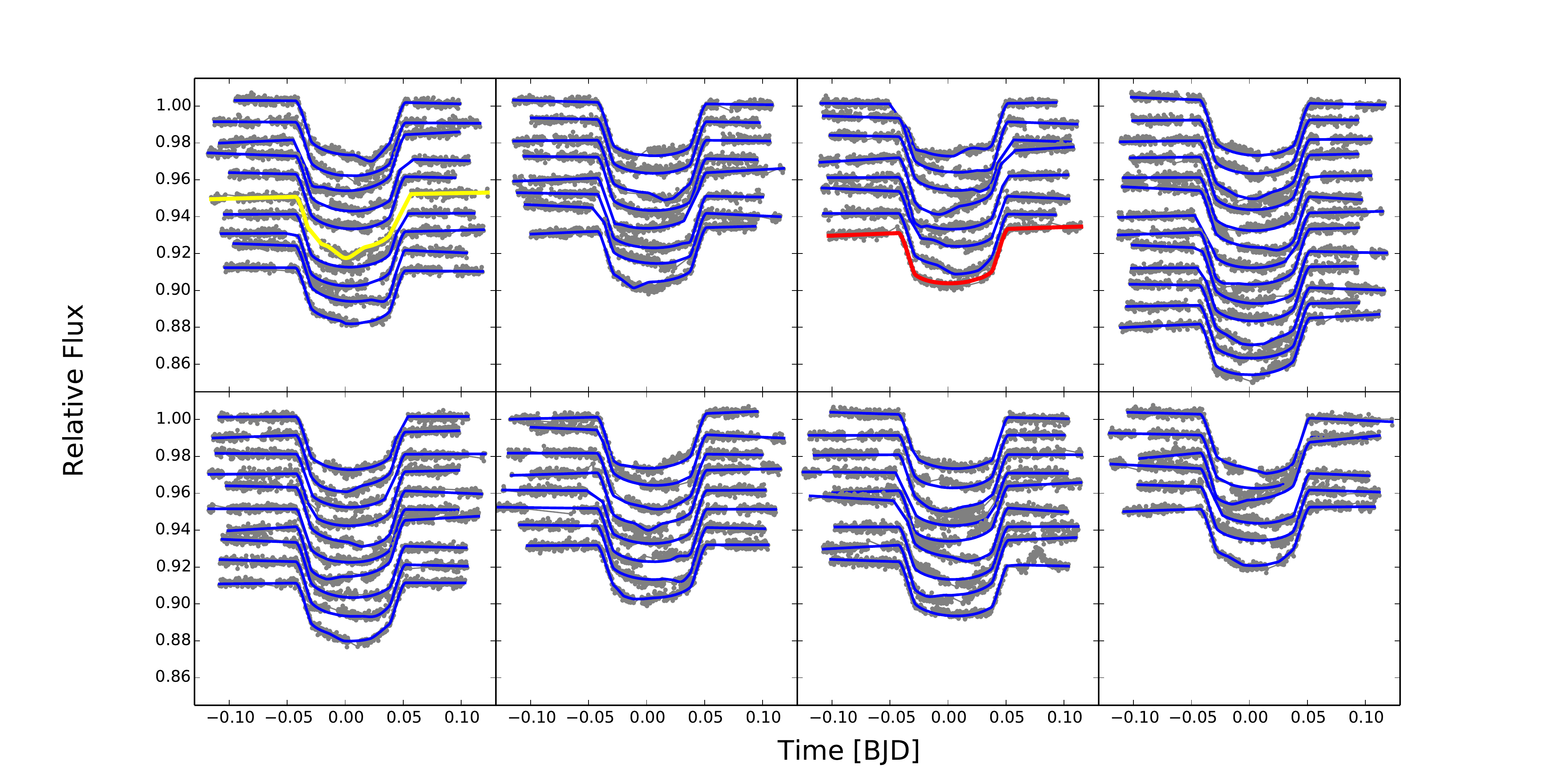}
\caption{Fit on the transits for every segment (from left to right, and top to bottom). The deepest transit, containing an occulted bright spot, is shown in yellow. A transit not affected by spots is in red.}
\label{transits_8segments}
\end{figure*}

\begin{figure*}[htbp]
\centering
\includegraphics[scale = 0.29]{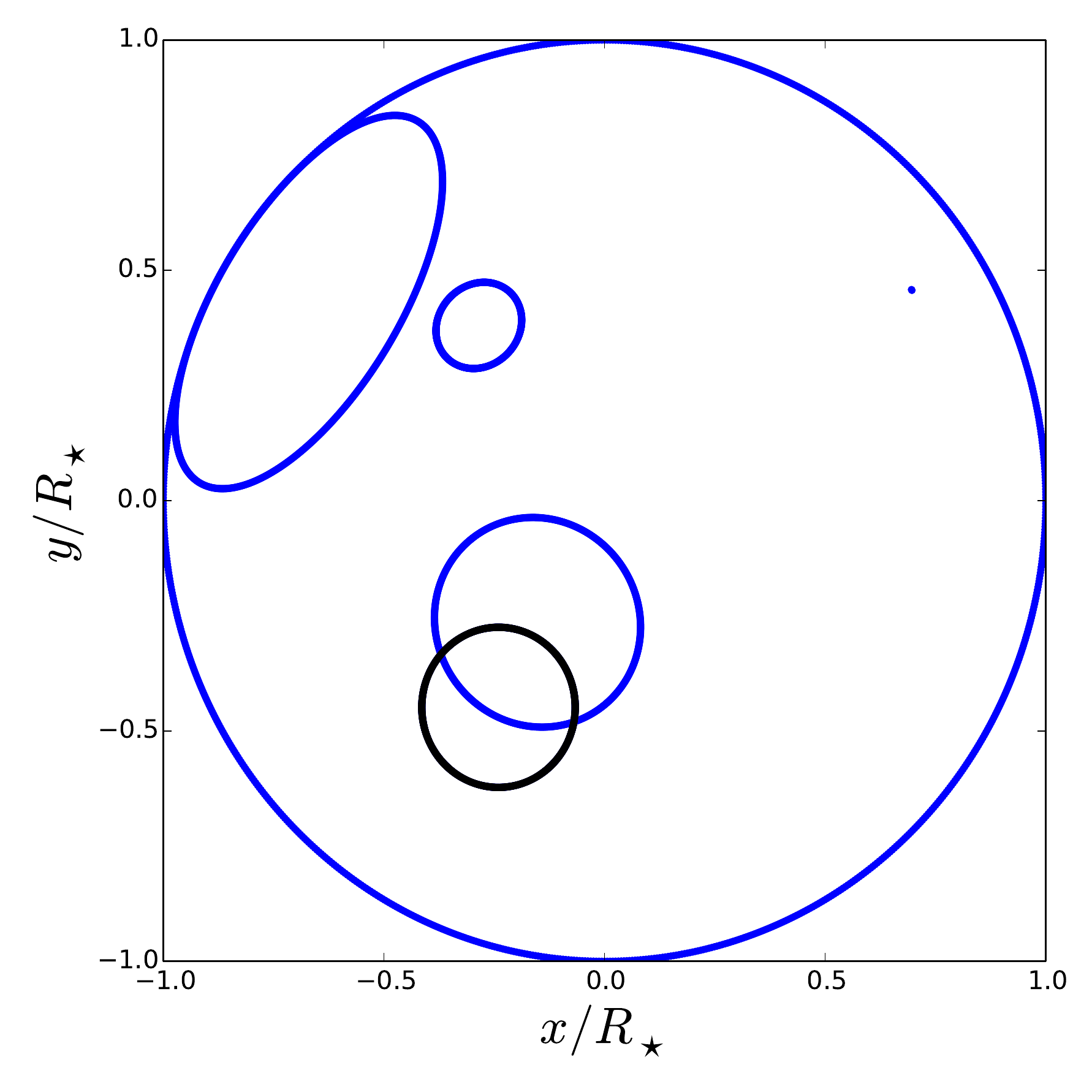}
\includegraphics[scale = 0.29]{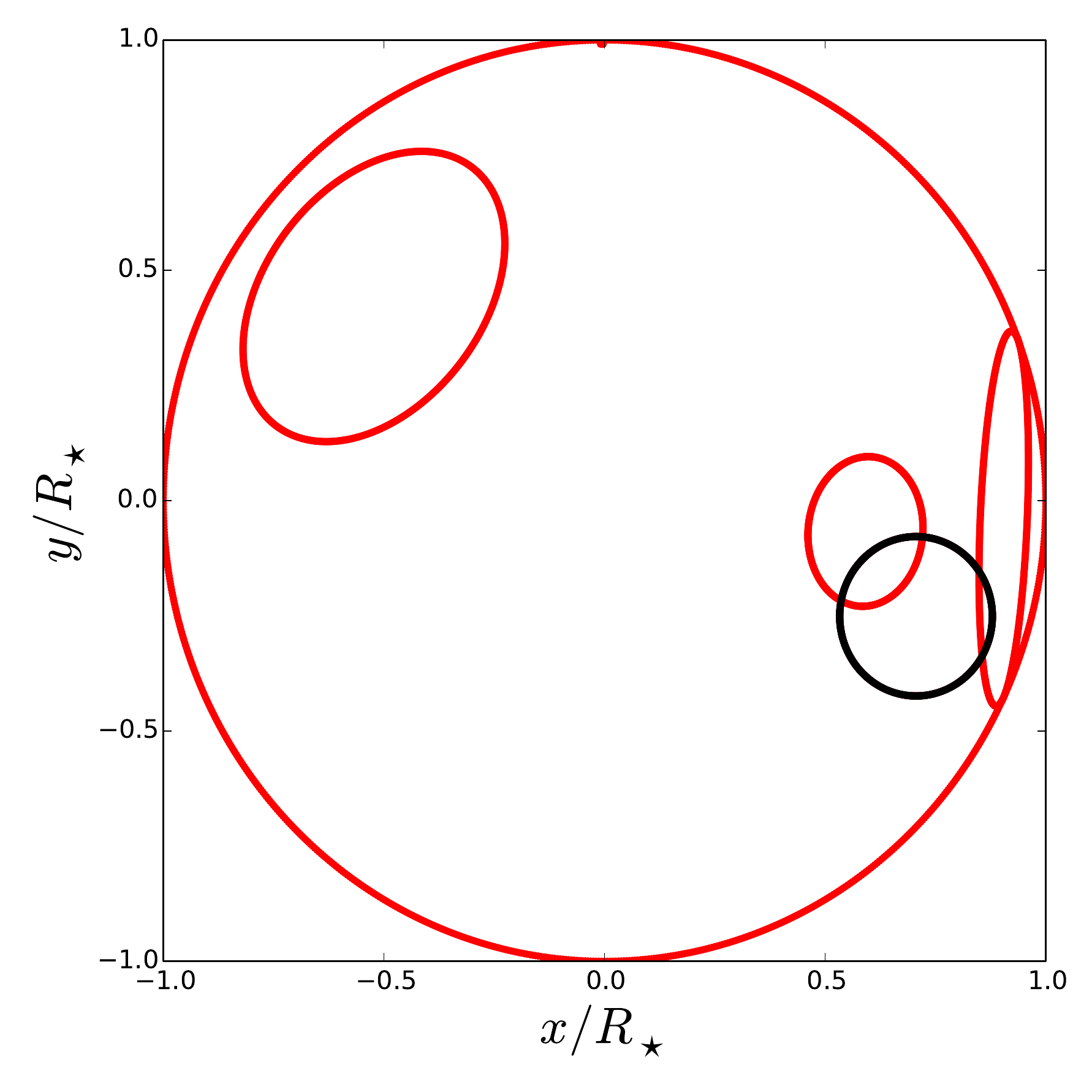}
\includegraphics[scale = 0.245]{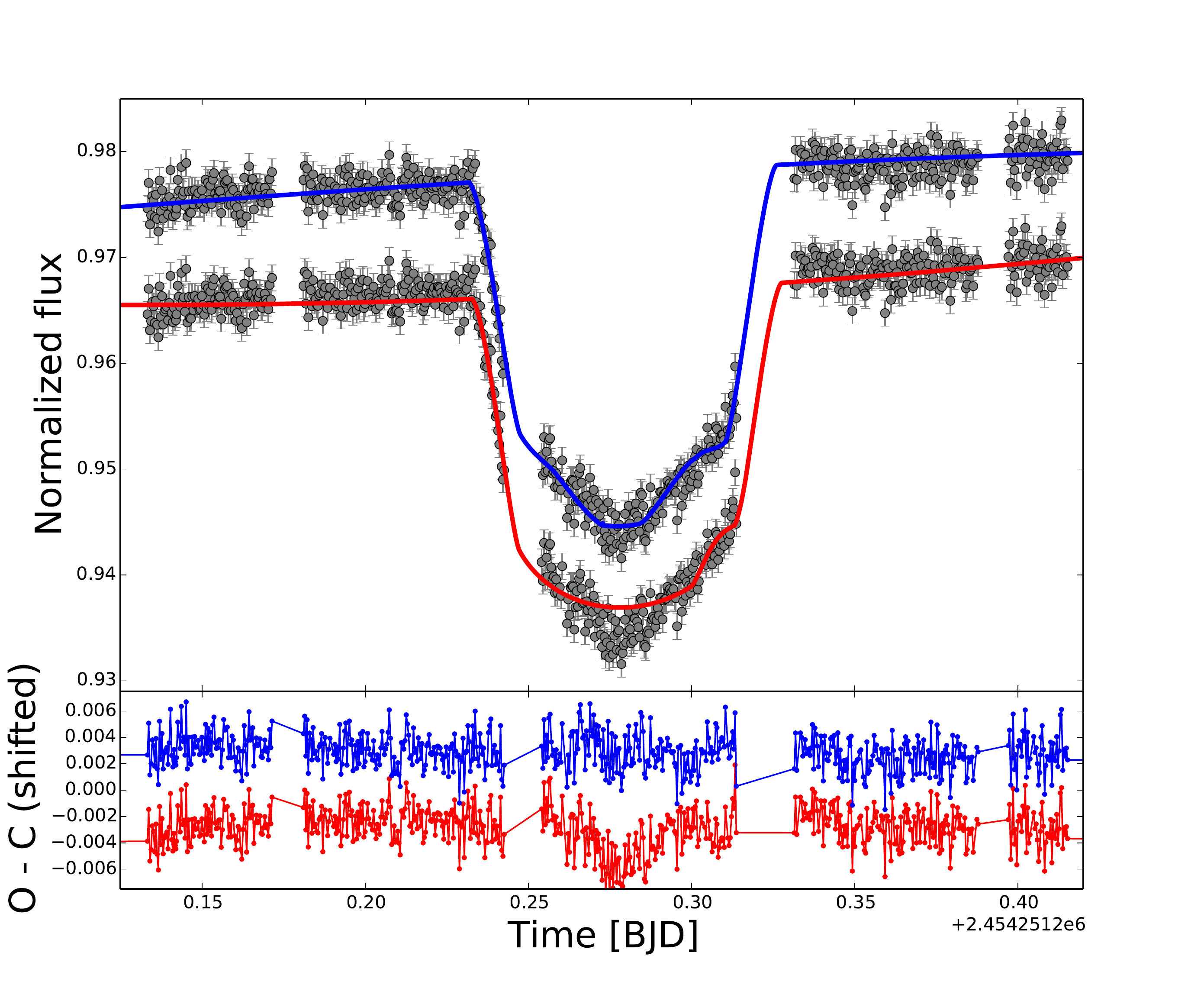}
\caption{\textit{Left:} Planet (black) and activity features (blue) configuration during the  deepest transit, with bright spots allowed. The bright spots are crossed by the planetary disc. \textit{Centre:}  Solution with three dark spots (red). \textit{Right:}  Deepest transit fitted with the dark-bright spot configuration (blue) and the three-dark spots model (red, shifted). The residuals are shifted for clarity and use the same colour code.}
\label{fitfacula6}
\end{figure*}

\section{Discussion}\label{disc}

The results of the modelling were used to explore the impact of the spots on the transit parameters.

\subsection{Least distorted transit}\label{individual}

According to \cite{czesla2009}, the deepest transits are less affected by occulted dark features than shallower ones. Therefore, their planet-to-star radius ratio should be closer to the true one. They thus interpolated a lower envelope to an average of the deepest transits of CoRoT-2 b. These were found to happen at the moments where the out-of-transit flux reaches the largest level. They fitted only $k_r$ and $i_p$, and obtained $k_r = 0.172 \pm 0.001$ (i.e. a 3\% larger $k_r$ than \citeauthor{alonso2009}) and $i_p = 87.7 \pm 0.2^\circ$. Their method assumes the dominance of dark spots over faculae, as found by \cite{lanza2009} by fitting the out-of-transit light curve.\\
With our approach, the assumption of the prevalence of dark spots inside transits could be checked. Even if we do not model faculae, we are still able to recover the need of features brighter than the stellar surface in the model. Such features would therefore increase the apparent transit depth.\\
We adopted a similar approach to \cite{czesla2009}, but worked on the deepest transit only in order not to average any activity feature. Figure \ref{transits_8segments} shows the series of transits with the best solution of fit $\mathcal{SPT}$ overplotted. The sixth transit in the first segment, in yellow in Figure \ref{transits_8segments}, is the deepest.\\
This transit was isolated from data set $\mathcal{SPT}$. According to fit $\mathcal{SPT}$, this transit is affected by a bright spot, whose position during the transit is shown on the left side of Figure \ref{fitfacula6}. 
By fitting a single transit, \ksint\ + \pastis\ does not disentangle the bright spot from the transit profile. For this reason, we fixed the configuration of the active regions obtained with fit $\mathcal{SPT}$ on the first segment of the light curve. Instead, the transit parameters $k_r$, $i_p$, $\rho_\star$, and the jitter were set as free parameters. Because of the low number of points in a single transit, the limb-darkening coefficients, the flux offset, and the contamination value were fixed.\\
The MCMC analysis yielded $k_r=0.1734^{+0.0010}_{-0.0016}$, $i_p = 86.02 \pm 0.27^\circ$, and $\rho_\star = 1.093 \pm 0.038\,  \rho_\odot$. The planet-to-star radius ratio is in $1\sigma$ agreement with the \cite{czesla2009} result. The low values of $i_p$ and $\rho_\star$, instead, have to be attributed to the distorted transit profile.\\ 
To check whether a dark spots-only solution can be found by our model, we fixed the planet-to-star radius ratio to the \citeauthor{czesla2009} value, and imposed three dark spots (i.e. with $f < 1$) for the fit of the deepest transit. A minimum of three spots was considered necessary. Indeed, two occulted dark spots at the borders of a transit can mimic a bright spot at the centre of the transit. A third non-occulted spot is needed to generate a possible out-of-transit flux variation if the other two are not sufficient. The contrast of the spots was fixed to the conservative solar value of 0.67 \citep{sofia1982}. The latitude of the occulted spots was fixed close to $-2.16^\circ$ in order to lie on the transit chord. Their longitudes were forced to lie in the visible stellar disk to help the fit. The latitude of the non-occulted spot was set to $30^\circ$.\\

The best dark spots-only configuration is plotted in the centre of Figure \ref{fitfacula6}. This solution and the one with a bright spot are compared on the right side of the figure. Without a bright spot, the distortions of the transit profile are not recovered. The Bayesian Information Criterion (BIC; \citealp{schwarz1978}) favours the model with a bright spot over the one with dark spots only: BIC$_{\mathrm{facula}}$ - BIC$_{\mathrm{dark \, spots}} \simeq -4$. This corresponds to a Bayes factor $\sim e^{-4} \simeq 0.018$ between the dark spots-only and the bright spot model.\\
This suggests that the unperturbed transit profile used by \cite{czesla2009} may actually be a lower envelope of the transits affected by bright spots or faculae, and may so lead to an overestimate of $k_r$. 
The transit profiles might be affected by bright spots, even if these spots have a negligible imprint on the out-of-transit part of the light curve. Including the transits in the modelling helps in recovering their presence.\\
Therefore, the unperturbed transit profile is more likely situated at a lower flux level than at the maximum level, which could indeed be produced by bright spots. Taking this into account, we inspected the modelled configuration of the system as in the previous cases, and we chose the twenty-sixth transit, in red in Figure \ref{transits_8segments}, as not affected by occulted activity features. We repeated the transit fit and obtained $k_r = 0.1689 \pm 0.0008$, $i = 87.63 \pm0.53^\circ$, and $\rho_\star = 1.336\pm0.062 \, \rho_\odot$, in agreement with the results obtained by fit $\mathcal{MN}$.\\

 Figure \ref{transits_8segments} also shows that for many transits, the model is not able to correctly reproduce the complex structure of the transit profile. This does not invalidate our previous results, as we used transits which were correctly modelled. However, the plots indicate the presence of short-lived or irregularly shaped features inside transits which are not reproduced by the model. The low level of detail introduced by the re-sampling on data set $\mathcal{SP}$ and the consequent approximated $\mathcal{SP}$ modelling hid the need of additional features for transit modelling.\\
\cite{silva-valio2011} needed seven to nine features to model each transit. In our approach, instead, this number of features is used to globally model tens of days of data.  
In order to get closer to the correct number of features needed for both the in-transit and the out-of-transit data, one possibility is to perform fit $\mathcal{SP}$ on data set $\mathcal{SPT}$ from the beginning, i.e. to use full resolution on the data. However, because of the large number of needed features and the large computational weight, we are not yet able to do this. We plan an optimization of the code in order to reduce its execution time and to be able to address the problem in a more efficient way.

\subsection{Impact of stellar activity on the transit parameters}\label{mostrel}

\begin{figure}[!htbp]
\centering
\includegraphics[scale = 0.5]{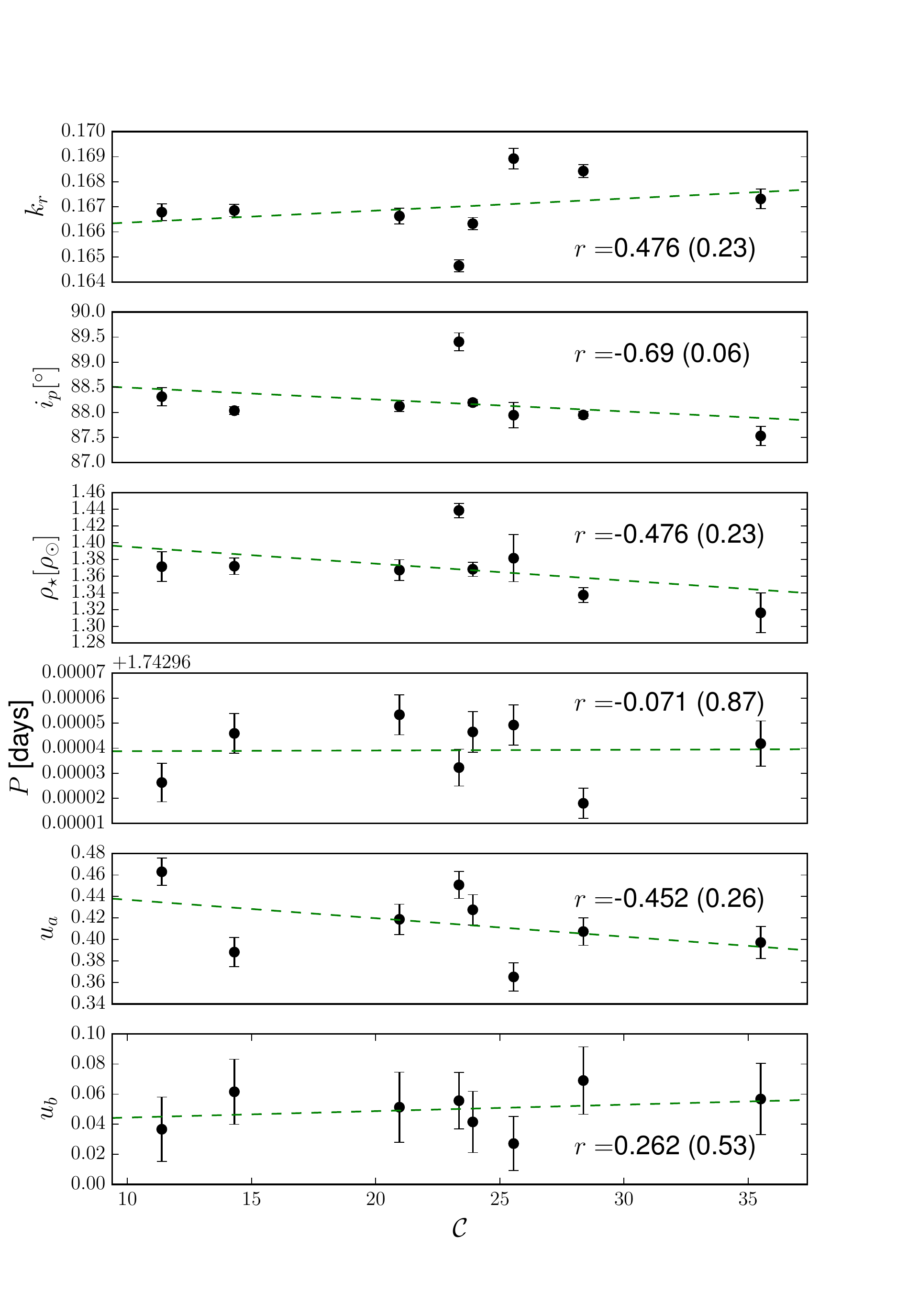}
\caption{Transit parameters as a function of the effective coverage factor, for all the segments. The values of the Spearman rank-order correlation coefficient are indicated. The corresponding p-values are in parentheses.}
\label{correlations}
\end{figure}

In Figure \ref{correlations}, all the transit parameters found with fit $\mathcal{SPT}$ are plotted as a function of the effective coverage factor $\mathcal{C}$ of the stellar surface, introduced in Section \ref{sppar}. In each panel, the Spearman rank-order correlation coefficient, followed by its corresponding two-sided p-value in parentheses, is shown. We conservatively adopted a significance level for the p-value of 0.05. Our results therefore show that the hypothesis of non-correlation with $\mathcal{C}$ is not rejected for all the transit parameters. We conclude that the observed scatter among the transit parameters is mainly due to statistical fluctuations caused by the division of the light curve in segments.\\
Several studies were presented where, without resorting to joint spot and transit modelling, the transit parameters are found to depend on the level of activity of the star. For example, a lower $\rho_\star$ attributed to starspots was observed for CoRoT-7 \citep{leger2009}, compared to the derived spectroscopic value. \cite{barros2014c7} later showed that this is mainly due to unresolved spot crossing events. Also, \cite{csizmadia2013} showed that limb-darkening coefficients vary with the fraction of the stellar surface covered by activity features. Our results point to the importance of a joint modelling in order not to incur such effects.

\subsection{Planet-to-star radius ratio and contamination}\label{disctr}
In our fit on locally normalized transits (fit $\mathcal{S}$), we used a contamination value of about 8.81\%. \cite{alonso2008} used the same normalization, but used a fixed contamination value of 5.6\%. They found $k_r \simeq 0.1667$, while fit $\mathcal{S}$ yields $k_r \simeq 0.1692$. A difference of $3\%$ in the contamination, therefore, produced an offset in $k_r$ of $1.5\%$. Instead, by keeping the same contamination value and a local (fit $\mathcal{S}$) or model-based normalization (fit $\mathcal{MN}$, $k_r \simeq 0.1682$) we found a difference of only $0.6\%$. The local and the \citeauthor{czesla2009}  normalization (fit $\mathcal{CZ}$), finally, yield a $1.7 \%$ difference. The choice of the normalization, therefore, may introduce a systematic error in the transit depth just as the choice of the contamination value.

\subsection{Planet radius}
The combination of fit $\mathcal{SP}$ and $\mathcal{SPT}$ allows, in principle, an unbiased  measure of the transit parameters. Indeed, it is not affected by the transit normalization, and takes the occulted activity features into account. However, our modelling approach is limited by the need to cut the light curve into segments, which introduces a scatter in the results because of the lower number of points to be fitted.\\ 
Fit $\mathcal{MN}$ can be adopted as a good compromise. Although it neglects occulted spots, this fit is not affected by the normalization or by the chop of the light curve into segments.
We therefore used $k_r$ and $\rho_\star$ obtained from fit $\mathcal{MN}$ to derive the radius of \corot-2 b. We used the Geneva stellar evolutionary tracks \citep{mowlavi2012} and the updated stellar atmospheric parameters of \cite{torres2012} (\teff$=5575\pm 70$ K, \feh$=-0.04 \pm 0.08$).\\
This yielded $R_P = 1.475 \pm 0.031$ \RJ. If, instead, we adopt the results of a standard normalization (fit $\mathcal{S}$), we obtain $R_P = 1.485 \pm 0.031$ \RJ, i.e. a less than 1\% larger radius. These values are about $\sim1\%$ different from the one found by \cite{alonso2008} ($1.465\pm0.029$ \RJ), but all of them are compatible. The inflated radius of the planet is confirmed.

\subsection{Limitations of the current model}
From the presentation of the results and their discussion, we identified some caveats which need to be addressed to improve the quality of our fits.\\
The first is the fixed number of activity features. An automatic incremental addition of the features could be implemented in order to minimize the residuals. A criterion should be chosen in order to stop the addition of features when residuals are smaller than a given threshold. This method would make the fit of short-lived features inside the transits easier. Second, the circular shape imposed on the features is another limitation of the model, and is linked to the previous one. The possibility of the features to overlap mitigates this problem, but many small features should overlap in order to model an active region with a complex shape.\\ 
Fitting $\lambda$ and \amax\ might not be the best way to propagate the uncertainties from fit $\mathcal{SP}$ to fit $\mathcal{SPT}$. However, if more parameters of the activity features are left free during fit $\mathcal{SPT}$, the chains do not converge.\\
We note that all features in our model obey the same limb-darkening law. An additional level of detail could be gained by modelling a limb-angle dependent contrast for bright spots, that is, by including the modelling of real faculae.\\
The lack of constraints on the features evolution parameters is another point which needs refinements in order to improve the realism of the model. Moreover, the analytic model of \ksint\ can be refined by the addition of differential rotation. \cite{frohlich2009} noticed that the information on differential rotation can be recovered only at the expense of some restrictions to the lifetimes of the features. However, if no differential rotation is modelled, two features at the same longitude but at different latitudes have a degenerate contribution to the total flux. Their contribution may therefore be equally modelled by a single, larger feature, which would simplify the convergence of the chains.\\  
Most of these problems cannot be efficiently explored before the computation time required by \pastis + \ksint\ is not importantly reduced. In fact, even if an analytic model was employed, some days of calculations were needed for the fit of each segment. This is due to the calculations required to model the overlap of active regions and transits, and to technical details of the code, which need to be optimized.

\section{Summary and conclusions}\label{conclusions}
We presented a method for the fit of transit photometry which takes the impact of activity features on the transit parameters into account. This approach is based on an improved version of the analytic code \ksint, which models both non-occulted and occulted features and their evolution. The method was applied to the light curve of \corot-2 in two ways. The first is based on the simultaneous modelling of activity features and transits in a non-normalized light curve. The second consists in normalizing the transits by using the modelling of the out-of-transit light curve, and then in a standard fit of the so-normalized transits. In particular, we allowed for the presence of bright spots in our model, which was not usually done in previous studies.\\
The results of our method were compared with other approaches presented in literature. We recovered the total effective coverage factor of the stellar surface all along the light curve. We found different rotational periods of the active regions at different longitudes, as observed by \cite{frohlich2009}.\\
We found that the choice of the normalization technique can introduce an offset in the transit depth of CoRoT-2 b as much as the choice of the contamination rate. Our results were then compared to those obtained with a standard local transit normalization, such as the technique used by \cite{alonso2008}. With a local normalization, and given the same contamination value, a $1.2\sigma$ increment of the measured planet-to-star radius ratio was found. Instead, the \cite{czesla2009} normalization, where bright spots or faculae are neglected, yielded a $2.3\sigma$ smaller transit depth. The stellar densities, instead, were found to be in agreement whether a model-based or a local normalization was performed, and at less than $2\sigma$ agreement if the \citeauthor{czesla2009} normalization was performed. We used the transit parameters obtained by our modelling to calculate the radius of CoRoT-2 b, and confirmed its inflated nature ($1.465\pm0.029$ \RJ).\\

Our analysis highlights the importance of stellar activity modelling during transit fit. Bright spots are particularly relevant in this respect. Including bright spots allowed us to correctly model the deepest transit profile without the need of increasing the transit depth, unlike  previous studies \citep{czesla2009,silva-valio2010}. Neglecting bright spots leads to the use of the lower envelope of the transits as an estimate of the unperturbed transit profile. We showed that the transit depth might be overestimated this way by almost 3\% and worsen the inflated radius issue. Instead, our method shows that the correct transit profile might be closer to the value given by an average run over all the transits. Other cases of planets transiting active stars need  to be analysed, however, in order to better constrain this result.\\

Thanks to the modelling of the time evolution of the features, it was possible to model longer parts of the light curve with respect to previous attempts presented in literature. Nevertheless, it was still necessary to cut the light curve in segments and to model each segment separately. A slight scatter was found among the transit parameters in the various segments. We showed that this has a likely statistical origin, given the small number of transits in each segment.
Developments that would allow for a more complex evolution of the activity features and the modelling of longer segments of the light curve would therefore remove the scatter among the transit parameters.\\

Improvements in the fitting techniques are also needed in order to model light curves affected by stellar activity spanning two or three years of observations, such as those which will be provided by PLATO 2.0 \citep{rauer2014}.\\
In order to develop our study in all these directions, we conclude by reaffirming the need of in-depth studies of other benchmark cases. 

\begin{acknowledgements}
S.C.C.B. acknowledges support by grants 98761 by CNES and the Funda\c c\~ao para a Ci\^encia e a Tecnologia  (FCT) through the Investigador FCT Contract No. IF/01312/2014. A.S. is supported by the European Union under a Marie Curie Intra-European Fellowship for Career Development with reference FP7-PEOPLE-2013-IEF, number 627202. Part of this work was supported by Funda\c{c}\~ao para a Ci\^encia e a Tecnologia, FCT, (ref. UID/FIS/04434/2013 and PTDC/FIS-AST/1526/2014) through national funds and by FEDER through COMPETE2020 (ref. POCI-01-0145-FEDER-007672 and POCI-01-0145-FEDER-016886). We thank S. Csizmadia and N. Espinoza for the fruitful discussions about limb darkening, D. Kipping for his help concerning the use of the \texttt{macula} code \citep{kipping2012}, R. Alonso for kindly providing us data to test our method, and R. F. D\'iaz for his explanations about Bayesian methods.
\end{acknowledgements}

\bibliographystyle{aa}
\bibliography{spots}

\appendix

\section{Additional tables}

\begin{table}[!htb]
\begin{center}
\caption{Main parameters of the CoRoT-2 system. From \cite{alonso2008}.}
\begin{tabular}{lrr}
\hline\hline
& Value& Error\\
Orbital period $P$ [d]&  1.7429964&0.0000017\\
Transit duration [d] & $\sim 0.09$ \\
planet-to-star radius ratio $k_r$ &  0.1667&0.0006\\
Orbital inclination $i_p$ [deg]& 87.84&0.10\\
Linear limb-darkening coefficient $u_a$ & 0.41&0.03\\
Quadratic limb-darkening coefficient $u_b$ & 0.06&0.03\\
Ratio of semi-major axis \\
to stellar radius $a/R_\star$ & 6.70&0.03 \\
Stellar density $\rho_\star$ [$\rho_\odot$] & 1.327& 0.018 \\
Eccentricity $e$ &  0 &(fixed)\\
Stellar mass $M_\star$ [$M_\odot$] & 0.97&0.06\\
Stellar radius $R_\star$ [$R_\odot$] & 0.902&0.018\\
Projected rotational velocity \vsini\ [km/s] & 11.85& 0.50\\
Effective temperature \teff\ [K] & 5625 & 120 \\
Planet mass $M_p$ [\MJ]  & 3.31&0.16\\
Planet radius $R_p$ [\RJ] & 1.465&0.029\\
\hline
\label{presc2}
\end{tabular}
\end{center}
\end{table}

\begin{landscape}
\begin{table}[thbp]
\begin{center}{
\caption{\label{tsegments} Transit parameters with their 68.3\% credible} intervals for each segment of fit $\mathcal{SPT}$; $t_i$ and $t_f$ stand for the initial and final times of each segment.}
\scalebox{0.8}{\begin{tabular}{ccccccccc}
\hline
\hline
\\
Segment & $t_i - 2454000$ [BJD]& $t_f - 2454000$ [BJD]& $k_r$ & $i_p [^\circ]$ &  $\rho_\star [\rho_\odot]$ & $P_{\mathrm{orb}}$ [days] & $u_a$ & $u_b$  \smallskip\\
\hline\\
1       &       242.009666      &       259.936338&     $0.16679\pm0.00033$     &       $88.31\pm0.18$  &       $1.371\pm0.018$ &       $1.742986\pm8\cdot 10^{-6}$        &       $0.463\pm0.013$ &       $0.037\pm0.021$ \\
2       &       260.019699      &       274.929539&     $0.16892\pm0.00042$     &       $87.94\pm0.25$  &       $1.381\pm0.028$ &       $1.743009\pm8\cdot 10^{-6}$        &       $0.365\pm0.013$ &       $0.027\pm0.018$ \\
3       &       275.011778      &       289.962123&     $0.16685\pm0.00024$     &       $88.03\pm0.08$  &       $1.372\pm0.01$  &       $1.743006\pm8\cdot 10^{-6}$        &       $0.388\pm0.014$ &       $0.062\pm0.022$ \\
4       &       290.025447      &       314.952613&     $0.16842\pm0.00025$     &       $87.95\pm0.07$  &       $1.337\pm0.009$ &       $1.742978\pm6\cdot 10^{-6}$        &       $0.407\pm0.013$ &       $0.069\pm0.022$ \\
5       &       315.032994      &       334.999139&     $0.16465\pm0.00024$     &       $89.41\pm0.18$  &       $1.439\pm0.009$ &       $1.742992\pm7\cdot 10^{-6}$        &       $0.451\pm0.013$ &       $0.056\pm0.019$ \\
6       &       335.00025       &       349.938381&     $0.16633\pm0.00025$     &       $88.2\pm0.06$   &       $1.368\pm0.009$ &       $1.743007\pm8\cdot 10^{-6}$        &       $0.428\pm0.014$ &       $0.042\pm0.02$ \\
7       &       350.013573      &       367.946176&     $0.16663\pm0.00032$     &       $88.12\pm0.11$  &       $1.367\pm0.013$ &       $1.743013\pm8\cdot 10^{-6}$        &       $0.419\pm0.014$ &       $0.051\pm0.023$ \\
8       &       368.022108      &       378.821162&     $0.16731\pm0.00039$     &       $87.53\pm0.19$  &       $1.316\pm0.024$ &       $1.743002\pm9\cdot 10^{-6}$        &       $0.397\pm0.015$ &       $0.057\pm0.024$ \\

\hline
\\
\end{tabular}}
\end{center}
\end{table}

\begin{table}[!bhtp]
\begin{center}{
\caption{\label{tsegmentsebop} Mean and standard deviations of the transit parameters on segments of the light curve with increasing length. The first column indicates the fitted fraction of the light curve; $t_i$ and $t_f$ stand for the initial and final times of each segment.}
\scalebox{0.8}{\begin{tabular}{ccccccccc}
\hline
\hline
\\
Fraction & $t_i - 2454000$ [BJD]& $t_f - 2454000$ [BJD]&  $k_r$ & $i_p [^\circ]$ &  $\rho_\star [\rho_\odot]$ & $P_{\mathrm{orb}}$ [days] & $u_a$ & $u_b$  \smallskip\\
\hline\\
0.125   &       242.618596      &       255.102084&     $0.16695\pm0.00089$     &       $89.41\pm0.39$  &       $1.433\pm0.017$ &       $1.74297\pm8\cdot 10^{-6}$        &       $0.497\pm0.014$ &       $0.024\pm0.022$ \\
0.25    &       242.618596      &       276.007638&     $0.16818\pm0.00085$     &       $88.57\pm0.29$  &       $1.412\pm0.021$ &       $1.74299\pm4\cdot 10^{-6}$        &       $0.436\pm0.014$ &       $0.036\pm0.03$ \\
0.33    &       242.618596      &       287.999862&     $0.16829\pm0.00086$     &       $88.4\pm0.23$   &       $1.403\pm0.019$ &       $1.742991\pm3\cdot 10^{-6}$        &       $0.411\pm0.015$ &       $0.044\pm0.033$ \\
0.5     &       242.618596      &       310.877715&     $0.16845\pm0.00077$     &       $88.65\pm0.21$  &       $1.422\pm0.015$ &       $1.742993\pm1\cdot 10^{-6}$        &       $0.428\pm0.008$ &       $0.02\pm0.017$ \\
0.66    &       242.618596      &       331.793366&     $0.1684\pm0.00079$      &       $88.43\pm0.17$  &       $1.401\pm0.014$ &       $1.742995\pm1\cdot 10^{-6}$        &       $0.419\pm0.01$  &       $0.028\pm0.022$ \\
0.75    &       242.618596      &       343.994937&     $0.16835\pm0.00079$     &       $88.32\pm0.15$  &       $1.393\pm0.014$ &       $1.742994\pm1\cdot 10^{-6}$        &       $0.429\pm0.007$ &       $0.014\pm0.013$ \\
1.0     &       242.618596      &       378.814125&     $0.16823\pm0.00075$     &       $88.29\pm0.15$  &       $1.39\pm0.013$  &       $1.742996\pm1\cdot 10^{-6}$        &       $0.427\pm0.009$ &       $0.025\pm0.019$ \\

\hline
\\
\end{tabular}}}
\end{center}
\end{table}
\end{landscape}

\end{document}